\documentclass[aps,prl,twocolumn,superscriptaddress,footinbib,floatfix,showpacs]{revtex4}

\usepackage[pdftex]{graphicx} 
\usepackage{dcolumn}
\usepackage{bm}

\newcommand{\be}{\begin{equation}}
\newcommand{\ee}{\end{equation}}
\newcommand{\ba}{\begin{eqnarray}}
\newcommand{\ea}{\end{eqnarray}}

\newcommand{\err}{\end{array}}
\newcommand{\bc}{\begin{center}}
\newcommand{\ec}{\end{center}}

\newcommand{\eg}{e.g.,~}
\newcommand{\ie}{i.e.,~}

\begin{document}
\preprint{LA-UR-07-XXXX}

\title[]{Tracing the Cosmic Web substructure with Lagrangian submanifold} 
\author{Sergei F. Shandarin}
\affiliation{Department of Physics and Astronomy, University of
  Kansas, Lawrence, KS 66045}
\author{Mikhail V. Medvedev}
\affiliation{Department of Physics and Astronomy, University of
  Kansas, Lawrence, KS 66045}
\affiliation{ITP, Natioal Research Center ``Kurchatov Institute", Moscow 123182, Russia}


\date{\today}

\begin{abstract}
A new computational paradigm for the analysis of substructure of the Cosmic Web in cosmological cold dark matter simulations is proposed. We introduce a new data-field --- the flip-flop field ---which carries wealth of information about the history and dynamics of the structure formation in the universe. The flip-flop field is an ordered data set in Lagrangian space representing the number of 
sign reversals of an elementary volume of each collisionless fluid element represented by a computational particle in a $N$-body simulation. This field is computed using the Lagrangian submanifold, \ie the three-dimensional dark matter sheet in the six-dimensional space formed by three Lagrangian and three Eulerian coordinates of the simulation particles. 
It is demonstrated that the very rich substructure of dark matter haloes and the void regions can be reliably and unambiguously recovered from the flip-flop field. 
\end{abstract} 


\maketitle


Modern redshift surveys such as 2dF Galaxy Redshift Survey \cite{2df} and the Sloan Digital Sky Survey \cite{sdss} and others reveal 
an intricate three dimensional structure in the spatial distribution of galaxies. Generic building blocks of the structure are:
haloes, filaments, walls, and voids. A useful abstraction helping to visualize this structure has been provided by the adhesion
approximation \cite{Gurbatov_etal:1989, Gurbatov_etal:2012, Hidding_etal:2012}. It is based on Burgers' model  of the equation of nonlinear 
diffusion \cite{Burgers:1974}. In the limit of infinitesimal viscosity it approximates the structure  by a  geometrical construction
 made of  two-dimensional curved faces, curvilinear edges at the surfaces crossings, and  zero-dimensional vertices
at the edge crossings. The vertices, edges and faces are analogous to haloes, filaments and walls respectively.
Both the observations and the model exhibit large empty regions called by astronomers as voids. 

Historically, haloes have attracted the most of attention in theoretical studies of the large-scale structure formation.  From the observational point of view,
haloes are most closely related to galaxies, galaxy groups and clusters of galaxies and they provide the bulk of
information about the structures in the universe. However, direct modeling of galaxy formation
based on fundamental laws of physics is precluded by enormous complexity of the physical processes involved, such as the highly nonlinear gravitational evolution of collisionless dark matter (DM) together with the 
hydrodynamical and thermal processes in baryons including star formation and the stellar wind feedback, shocks and supernovae explosions, gas accretion onto black holes in active galactic nuclei and the feedback via relativistic jets, and others.
Hence various semi-empirical models of galaxy formation have been suggested, see \eg \cite{Angulo_etal:2013} and
references therein. In particular, it has been argued that galaxies
are formed in the host DM haloes of corresponding masses. The DM haloes themselves  are formed in a
chain of mergers of smaller DM haloes which may start from tiny 
haloes of a planet
mass
~\cite{Diemand_etal:2005}. When two or more haloes merge 
their remnants may survive for a long time as subhaloes and/or streams within the resultant halo. 
Therefore, DM haloes are likely to have a hierarchical  structure  resembling a Russian doll or `matryoshka',
where each  subhalo includes a number of  even smaller subhaloes down to the smallest haloes allowed by the initial power spectrum \cite{Diemand_etal:2005,Ghigna_etal:1998}. 

\begin{figure}
\includegraphics[scale=0.7]{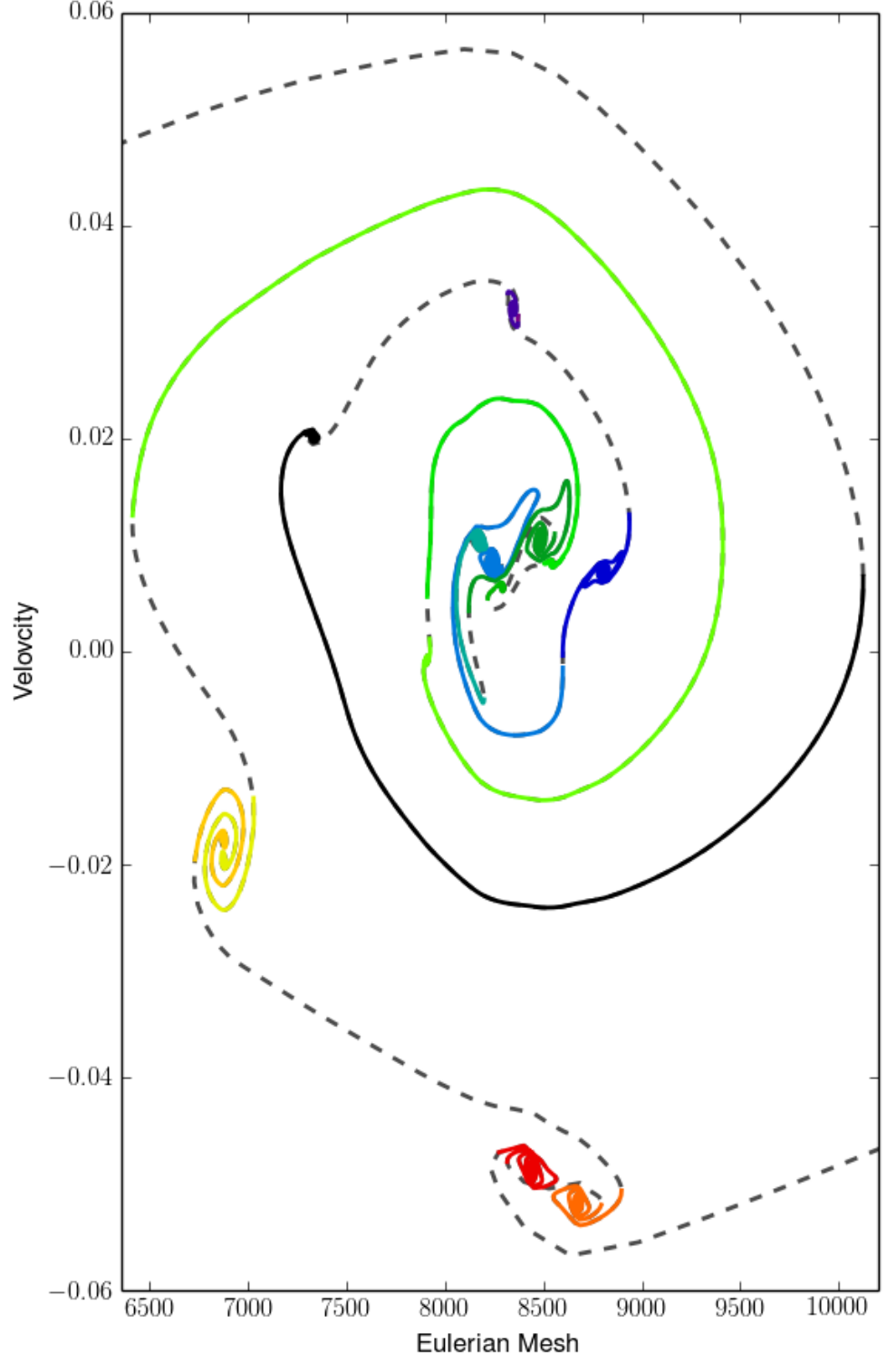}
\caption{\label{fig:1d_phase_space} The phase space of a one-dimensional halo simulated from random but smooth initial
condition. The individual subhaloes are shown by different colors}
\end{figure}

In early cosmological $N$-body simulations the haloes were loosely defined as compact concentrations of the simulation
particles in configuration space. A particularly popular simple technique used for this purpose and called 
`friends of friends' (FOF) algorithm was adopted from percolation analysis \cite{Shandarin:1983, Davis_etal:1985}.
According to this method one firstly finds all `friends' of each particle by linking every particle in the simulation 
with all neighbors  separated by less than a chosen distance -- the linking length. 
Then applying the criterion: a friend of my friend is my friend,  one can identify all groups of particles consisting of  friends. 
Choosing a particular value of the linking length  (often $\sim20\%$ of the mean particle separation)
one can select a particular set of groups and call them haloes. 
Other more sophisticated methods that  identify
both haloes and subhaloes, some of which used only the positions of particle other also  the phase space information, 
 have been suggested as well, see \eg  \cite{Vogelsberger_etal:2011,Knebe_etal:2013, 
 Hoffmann_etal:2014} and references therein. The methods using only the configuration space information 
 suffer from projection effect that causes dynamically distinct structures in phase space to overlap 
 in configuration space (for illustration see  \eg fig. \ref{fig:1d_phase_space}).  Using all dynamical
 information provided by phase space is complicated by the fact that it is not a metric space
 \cite{Ascasibar_Binney:2005}.   
 
In this letter we propose a novel technique which allows one to identify haloes and subhaloes by analyzing the mapping ${\bf x} = {\bf x}({\bf q},t)$ where ${\bf x}$ and {\bf q} 
are the coordinates of the particles in Eulerian and Lagrangian spaces respectively.  Topologically, this mapping, referred to as the Lagrangian submanifold,  is a three-dimensional sheet in the the six-dimensional $({\bf q, x})$ space. 
The method is based on a concept of a DM sheet in phase space  
 ${\bf v} = {\bf v}({\bf x},t)$   successfully employed to improve accuracy of the estimates of the density, velocity and other parameters in standard cosmological $N$-body simulations
\cite{Shandarin_etal:2012, Abel_etal:2012}. 
The major difference between this concept and the conventional one is in the different interpretation of the role of the particles 
in the simulations of the evolution of the continuous DM medium. 
Instead of the common interpretation of particles as carriers of mass, it was suggested to treat them as massless markers of the vertices in a tessellation of the three-dimensional DM sheet 
in six-dimentional phase space. 
The particles' mass is uniformly distributed inside each tetrahedra of the tessellation \cite{Shandarin_etal:2012, Abel_etal:2012}.
Once the tessellation is built in the initial state of the simulation, it must remain intact 
through the whole evolution because of the Liouville's theorem, as long as the thermal velocities of the DM particles are vanishing. This requirement results in a significant difference between this approach and
Delaunay tessellation suggested in \cite{Scharp_vdW:2000} for estimating the density from
particle distributions.

\begin{figure}
\includegraphics[scale=0.7]{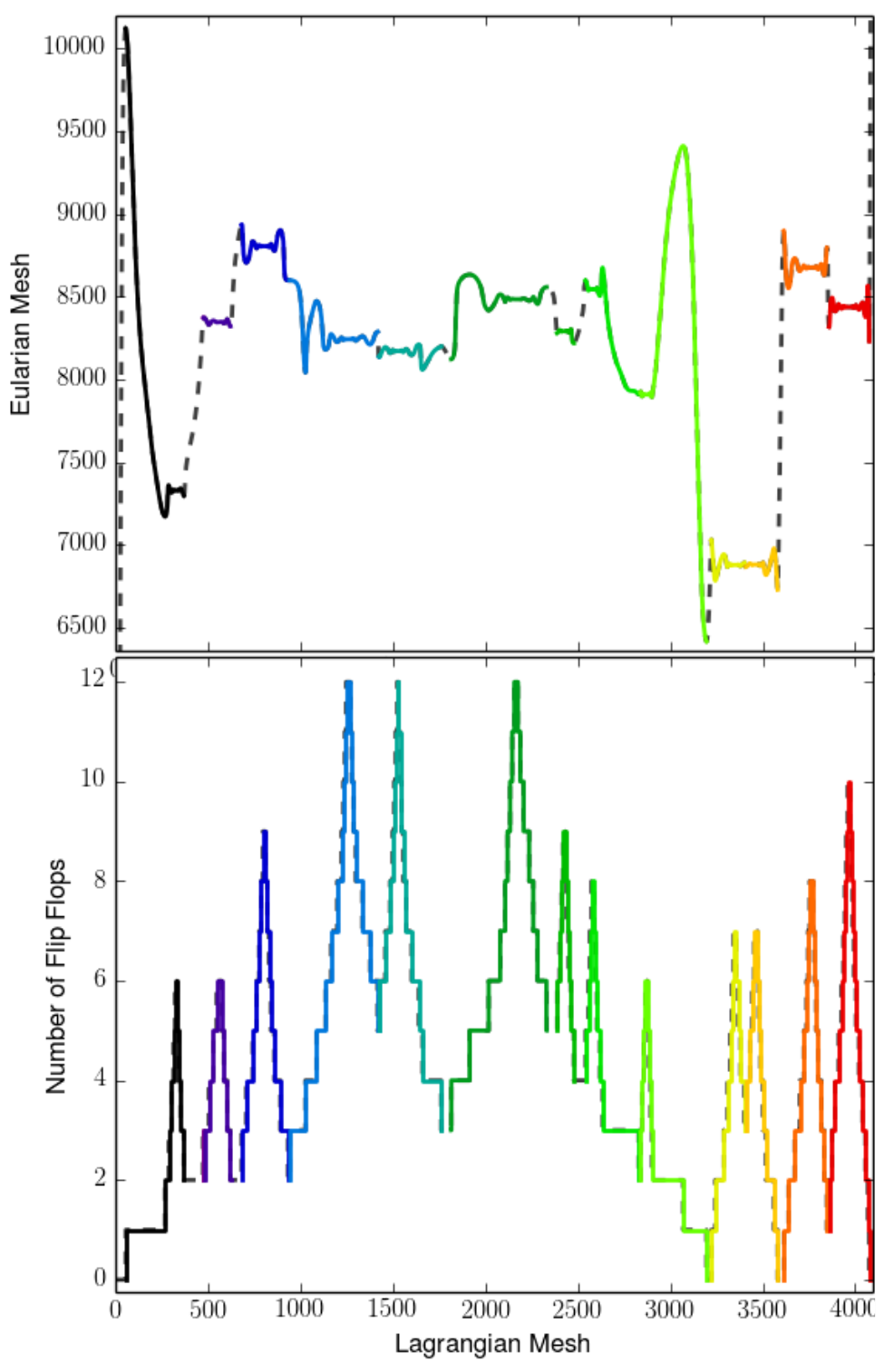}
\caption{\label{fig:1d_lm_nff}
Fields  $x(q)$ and  $n_{\rm ff}(q)$ are plotted 
in the top and bottom panels respectively for the halo shown in fig. \ref{fig:1d_phase_space}.}
\end{figure}

The particles being the vertices of the tetrahedra  describe all deformations occurred to the 
geometry of the tessellation. However it remains  continuous in both six-dimensional phase
space (${\bf x}, {\bf v}$) and in (${\bf q}, {\bf x}$) space. 
In particular, the variations of tetrahedra volumes result in the corresponding change of the tetrahedra 
densities.  This property is especially valuable because it makes the tessellation self-adaptive to 
the growth of density perturbations with time. 
We stress that whereas both (${\bf x}, {\bf v}$) and (${\bf q}, {\bf x}$) spaces contain all the information about a dynamical system, the latter is a {\it metric} space and hence superior to the non-metric phase space. Moreover, the Lagrangian submanifold mapping, ${\bf q=q(x)}$, is a single-valued function, unlike the phase-space mappings ${\bf v=v(x)}$ or  ${\bf x=x(v)}$ which are multivalued.

We now illustrate the main idea of the proposed Lagrangian submanifold technique with a halo formed in
one-dimensional $N$-body simulation of a collisionless cold DM medium in an expanding universe.

Figure \ref{fig:1d_phase_space} shows the phase space of the  halo evolved in the
universe from smooth random initial condition. 
The halo can be naturally defined as the region in Eulerian space 
where the number of streams is greater than one. The number of stream changes by two at caustics 
where the tangent to the phase space curve becomes
vertical and the density in the corresponding stream  becomes formally infinite.
One can see a complicated substructure that consists of a number of subhaloes and streams
shown by different colors.  
It is obvious from the figure that identifying individual subhaloes in the configuration (Eulerian) space 
is difficult even in a simple one-dimensional model due to projection effects and the presence of tidal streams.
It becomes even more challenging in three-dimensional simulations, see \eg  \cite{Knebe_etal:2013, 
Hoffmann_etal:2014} and reference therein. 

Let us follow along the phase space curve in fig. \ref{fig:1d_phase_space} starting from the top point of the spiral on the left boundary of the box through the bottom point on the right boundary of the box. Along this path, the initial (Lagrangian) coordinates $q_{\rm i}$ of the particles, which are in essence their IDs, increase 
monotonically while their final (Eulerian) coordinates $x_{\rm i}$ are not monotonic 
as is also seen in the top panel of fig. \ref{fig:1d_lm_nff}.  In other words there are fluid elements with $x_{\rm i+1} < x_{\rm i}$ while $q_{\rm i+1} > q_{\rm i}$. 
We will dub every swap of the coordinates of the two neighboring particles on the curve as a flip-flop. 
The analog of this phenomenon in a multi-dimensional space is a formal change of the sign
of the volume of a fluid particle when it turns inside out. 
The total number of flip-flops experienced by every fluid
particle is shown in the bottom panel of fig \ref{fig:1d_lm_nff}.  
Colors show individual peaks of the flip-flop field in Lagrangian coordinates. 
The correspondence of the flip-flop peaks in Lagrangian space to the individual subhaloes
in the phase space is remarkable. Note that the tidal streams and their progenitor halos are also easily, unambiguously and robustly identified via the flip-flop field, cf. the bottom panel of figure  \ref{fig:1d_lm_nff} and the phase-space figure \ref{fig:1d_phase_space}.

 
Next we show that the flip-flop field $n_{\rm ff}({\bf q})$  exhibits  similar features in 
a generic three-dimensional $N$-body simulation. The Lagrangian submanifold technique was implemented in the publicly available cosmological TreePM/SPH code GADGET \cite{Springel:2005} to compute the flip-flop field. The flop-flop module works as follows. At each time step and for each particle, the Jacobian $J({\bf q},t) = |\partial x_i/\partial q_j|$ is evaluated and compared it with its value at the previous time step. If the sign of the Jacobian changes, the number of flip-flops for this particle is increased by one. 

Initial conditions were generated with N-GenIC code with the standard $\Lambda$CDM cosmology, $\Omega_m=0.3,\Omega_\Lambda=0.7, \Omega_b=0, \sigma_8=0.9, h=0.7$ and the initial redshift $z=50$. A set of simulations were carried out with boxes ranging from $100 h^{-1}$ to $1 h^{-1}$~Mpc. For illustration purposes, we present here a relatively small zoomed-in simulation with $256^3$ DM particles in a box with the comoving size of $1 h^{-1}$~Mpc with the force resolution of $0.75 h^{-1}$~kpc. The chosen size of the box is obviously too small for the purpose of deriving statistically valid properties of the haloes. However the main purpose of this work is different, namely to demonstrate that the flip-flop field of haloes in a highly nonlinear dynamic  state  retains rich information about the substructure in haloes.

\begin{figure}
	\centering
	\centerline{\includegraphics[scale=0.75]{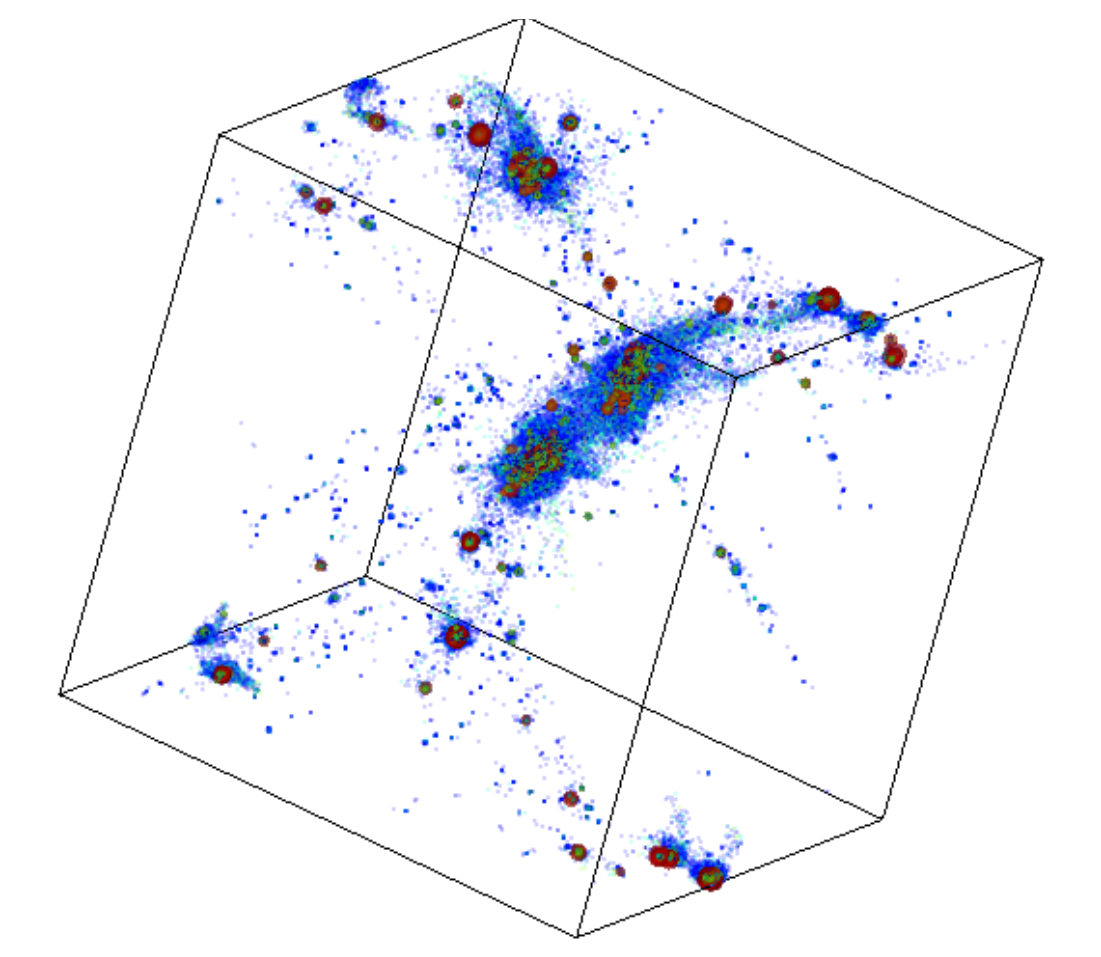}}
	\caption{\label{fig:3d_E} The dot plot of the structure in the simulation of $1/h$~Mpc box in the $\Lambda$CDM cosmology 
	at  $z=0$.  The sizes and colors from blue to red corresponds  
	to the range $n_{\rm ff} \ge 6$. }
\end{figure} 
\begin{figure}
	\centering
	\centerline{\includegraphics[scale=0.75]{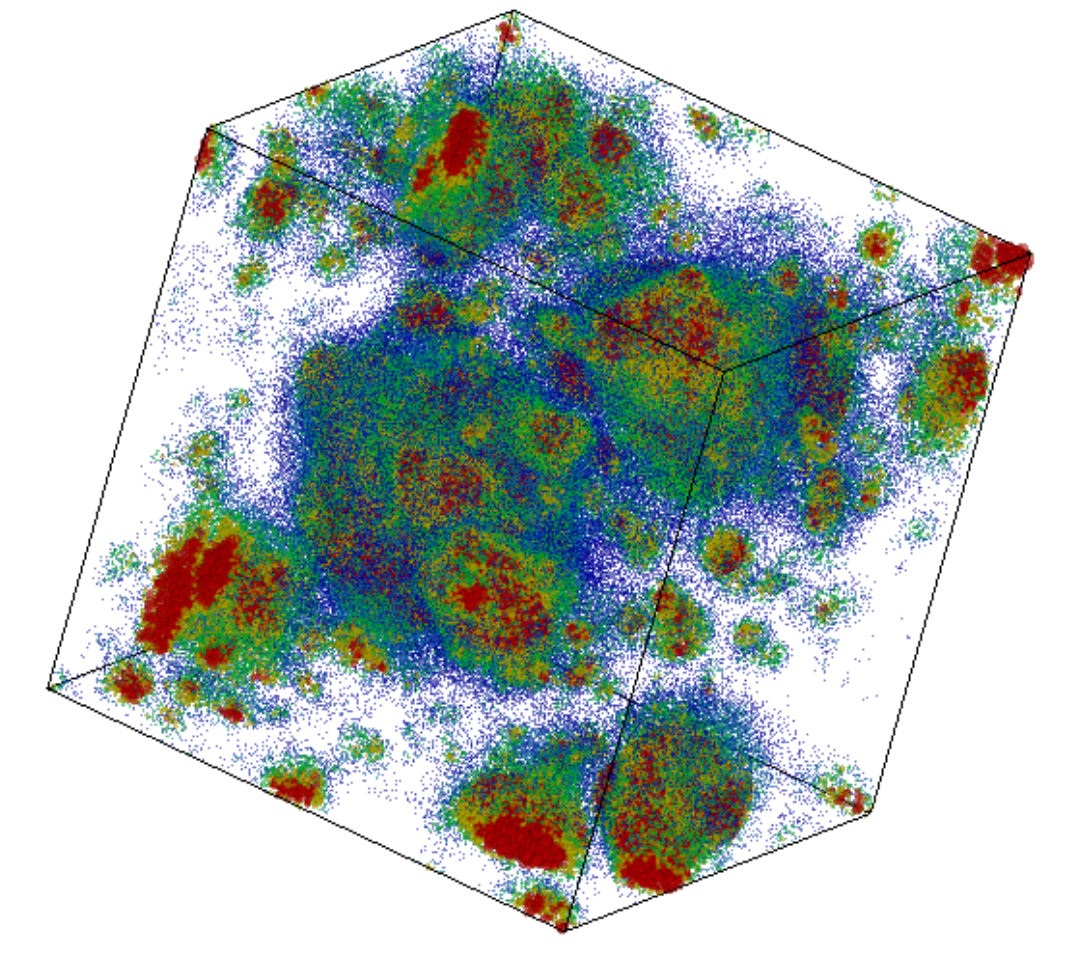}}
	\caption{\label{fig:3d_nff} The dot plot of the flip-flop field in Lagrangian space. The particles sizes and colors are similarl to that in fig. \ref{fig:3d_nff}.} 
\end{figure} 

Figure \ref{fig:3d_E} shows the flip-flop field in Eulerian space with $n_{\rm ff}$ 
from 6 to 420, the maximum number of flip-flops in this simulation. Obviously, this flip-flop field traces the distribution of matter in the universe. 
The sizes and colors (from blue to red) of the particles represent the number of flip-flops
 \footnote{Boosting the sizes of less abundant particles with high flip-flop numbers allows on to see them in a crowd of much more numerous particles with low flip-flop numbers.}.
 A dedicated analysis shows that the clumps of red particles (i.e., those with large flip-flop numbers) are not individual subhalos, but instead belong to different subhalos in the Lagrangian space,  fig. \ref{fig:3d_nff}. This indicates the inability of a configuration space-based analysis to disentangle all the substructure.

\begin{figure}
\includegraphics[scale=0.4]{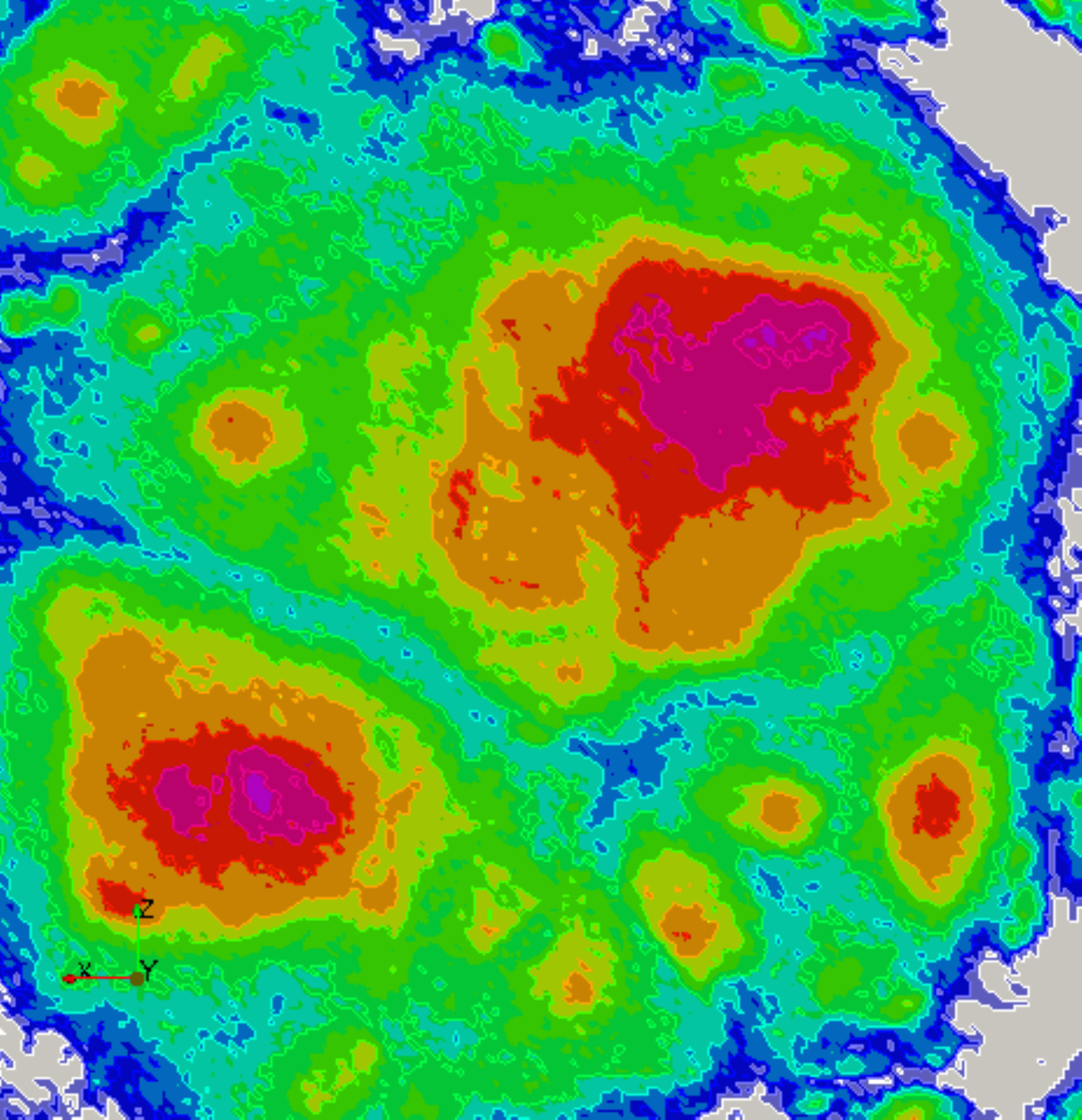}
\caption{\label{fig:slice} The contour plot of $n_{\rm ff}(q)$  field on a 2D cross-section plane through the center of the cube  
shown in fig. \ref{fig:3d_nff}.
	Contours from gray to magenta are: $n_{\rm ff} =$ 1.5, 2.5, 4, 7, 12, 27, 45, 75, 120, and 175. }
\label{fig:contour_plot}
\end{figure}

Figure \ref{fig:3d_nff} shows the corresponding flip-flop field in Lagrangian space with the same color coding. 
One can clearly see that they form a large number of distinct flip-flop peaks in Lagrangian space.
In order to reveal the much greater richness and complexity of the structure of
subhaloes in the flip-flop field, we also plot a two-dimensional slice through Lagrangian space
in fig. \ref{fig:slice}. This figure shows a complex hierarchy of peaks in greater detail.
We also stress that the regions with zero flip-flops are, by definition, voids. Thus, the flip-flop formalism is a superior void finder, as it does not suffer from the poor density contrast or other issues. 

\begin{figure}
\includegraphics[scale=0.45]{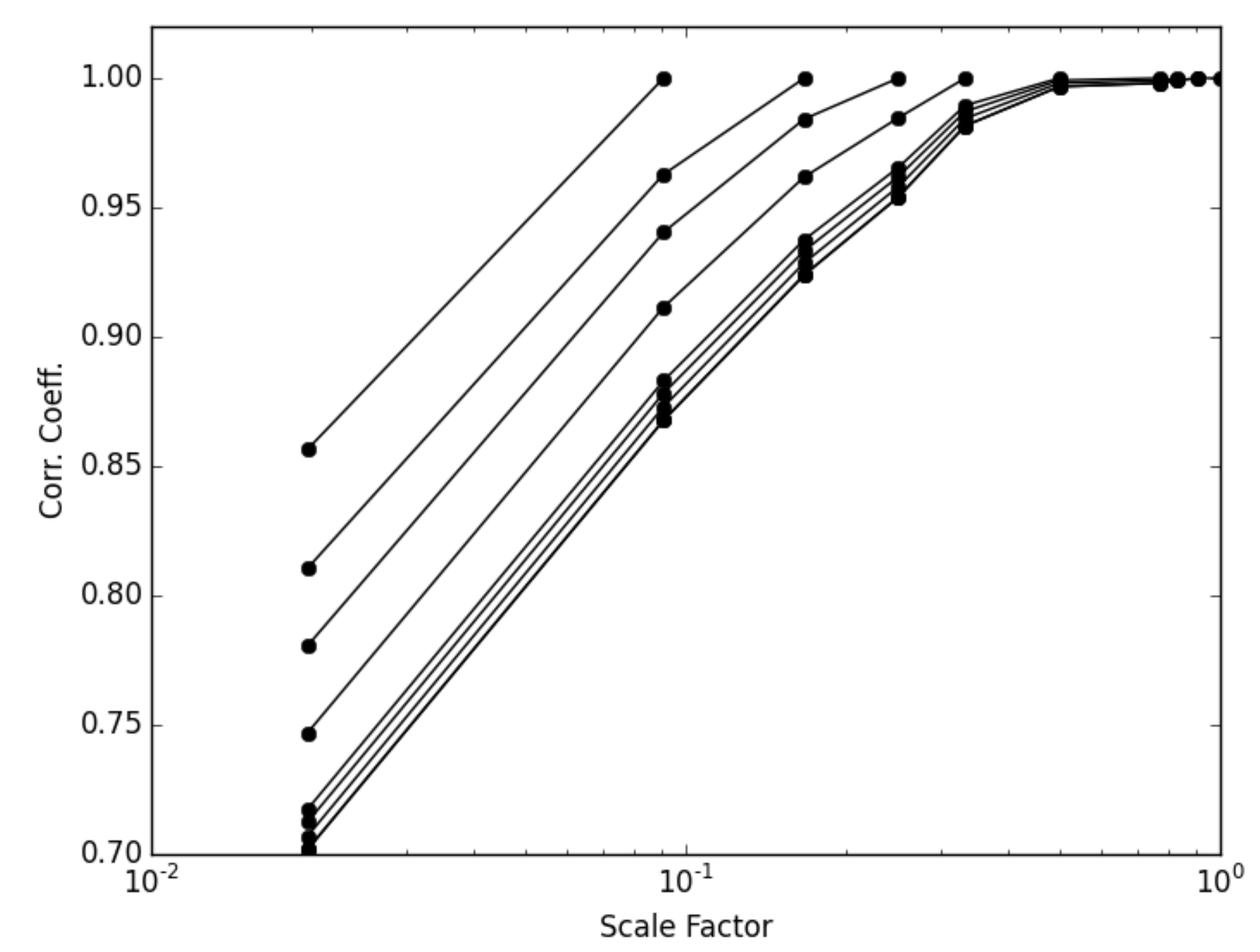}
\caption{\label{fig:corr_coef} The correlation coefficient between the flip-flop fields at different epochs.
Every flip-flop field is correlated with several previous nonlinear stages. 
The top point on each curve marks  the correlation coefficient of the field with itself and thus is 
exactly unity.}

\end{figure}

A similar analysis as a function of redshift suggests that there are at least two
distinct stages in the evolution of the flip-flop field: `fast and early' and `slow and late'. 
Indeed, starting from the onset of nonlinearity, the flip-flop field evolves rapidly at $z\gtrsim1$ and slows down afterwards. 
The similarity and difference of the flip-flop field evaluated at different redshifts can be 
quantified by computing 
the correlation coefficient between the fields at two different epochs. Figure \ref{fig:corr_coef} shows 
the correlation coefficient for several pairs of the flip-flop fields. The lower line shows the correlation coefficient 
between the field at $a=1~(z=0)$ and the fields at several previous stages. 
The top point on every curve shows the correlation coefficient  of the field with itself and thus
its value is one. The correlation coefficient monotonically decreases  with the separation between the epochs.
At first five stages shown,  it steadily decreases which means that new peaks in the flip-flop field keep forming
after the previous epoch.  In contrast, the bottom four curves `pile up', which suggests that the  evolution  slows down significantly. 

Summarizing the results we conclude that the flip-flop field carries wealth of information 
about the substructures in the Cosmic Web in the form of peaks separated by the valleys 
with lower counts of  flip-flops, as is illustrated in fig. \ref{fig:slice}.
The peaks often consists of several higher peaks which in turn may consist of even a higher peak,
forming a nesting structure resembling a Russian doll or `matryoshka'-doll. The topography of the 
flip-flop landscape  evolves rapidly after the onset of nonlinearity marked by the origin of
the first regions with $n_{\rm ff} > 0$. Then its evolution seems to freeze or considerably 
slow down (see fig. \ref{fig:corr_coef})  despite  the peak heights continue to grow, which indicates ongoing rapid dynamics inside the halos themselves. Qualitatively
similar pattern is also observed in two-dimensional case (not discussed here). We believe that
the suggested method represents a valuable addition to the suite of various techniques suggested
for studies of substructures in the Cosmic Web, see \cite{Vogelsberger_etal:2011,Knebe_etal:2013,Hoffmann_etal:2014} and references therein.

SSh acknowledges the support  by the Templeton Foundation and  sabbatical support at Kapteyn Astronomical Institute at the University of Groningen The Netherlands and by Argonne  National Labs where the significant part of the work was done. SSh also thanks S. Habib for useful discussions.
MM acknowledges partial support by DOE and NSF via grants DE-FG02-07ER54940 and AST-1209665

\bibliographystyle{prsty}  
\bibliography{mybibliography}

\end{document}